\documentclass[aps,prl,amsmath,amssymb,twocolumn]{revtex4}

\usepackage{graphicx}
\usepackage{dcolumn}

\begin{document}


\title{Zeno and anti--Zeno effects for quantum Brownian motion}


\author{Sabrina Maniscalco}
\affiliation{Department of Physics, University of Turku, Turun
yliopisto, FIN-20014 Turku, Finland}
\author{Jyrki Piilo}
\affiliation{Department of Physics, University of Turku, Turun
yliopisto, FIN-20014 Turku, Finland}
\author{Kalle-Antti Suominen}
\affiliation{Department of Physics, University of Turku, Turun
yliopisto, FIN-20014 Turku, Finland}

\email{sabrina.maniscalco@utu.fi}

\date{\today}

\begin{abstract}
In this paper we investigate the occurrence of the Zeno and
anti-Zeno effects for quantum Brownian motion. We single out the
parameters of both the system and the reservoir governing the
crossover between Zeno and anti-Zeno dynamics. We demonstrate
that, for high reservoir temperatures, the short time behaviour of
environment induced decoherence is the ultimate responsible for
the occurrence of either the Zeno or the anti-Zeno effect. Finally
we suggest a way to manipulate the decay rate of the system and to
observe a controlled continuous passage from decay suppression to
decay acceleration using engineered reservoirs in the trapped ion
context .
\end{abstract}

\maketitle

The quantum Zeno effect (QZE) predicts that the decay of an
unstable system can be slowed down by measuring the system
frequently enough \cite{Misra}.  In some systems, however, an
enhancement of the decay due to frequent measurements, namely the
anti-Zeno or inverse Zeno effect (AZE), may occur \cite{Lane}.

In this paper we focus on the quantum Brownian motion (QBM) model
\cite{HuPazZhang,Maniscalco04b,Weissbook} which is a paradigmatic
model of the theory of open quantum systems. This model, dealing
with the linear interaction of a particle with a bosonic reservoir
in thermal equilibrium, is widely used in several physical
contexts. It describes the dynamics of a particle interacting with
a quantum field in the dipole approximation \cite{Zurek}, as well
as a quantum electromagnetic field propagating in a linear
dielectric media \cite{Anglin96}. The model is used in nuclear
physics to describe, e.g., the two-body decay of an unstable
particle \cite{Joichi}. In quantum chemistry, the QBM model
describes the quantum Kramers turnover, which forms the basis of
modern theory of activated processes \cite{HanggiRev}.

Recently, this model has been investigated to explain the loss of
quantum coherence (decoherence) due to the interaction between the
system and its surroundings. In particular, the absence of
macroscopic quantum superpositions in the classical world has been
explained, using the QBM model, in terms of environment induced
decoherence (EID) \cite{Zurek}. With the last term we mean here a
process which transforms a highly delocalized state in position
and/or momentum, e.g. a superposition of coherent states, into a
localized classical state.

To the best of our knowledge, the conditions for the occurrence of
the quantum Zeno and anti-Zeno effects have never been considered
for the QBM model. The aim of this paper is to investigate the
Zeno and anti-Zeno phenomena in a system, namely the damped
harmonic oscillator, which possesses a classical limit and where
it is therefore possible to monitor the transition from quantum to
classical dynamics caused by decoherence induced by the
environment. Previous studies focus on few level systems and deal
with completely quantum states, such as the spin, which have no
classical analogue \cite{FacchiRev,Kurizki05}.

Our main result is the demonstration that the occurrence of either
the Zeno or the anti-Zeno effect stems from the short time
behaviour of the environment induced decoherence, which therefore
drives the Zeno--anti-Zeno crossover. On the other hand, one can
use the QZE or the AZE to manipulate the quantum-classical border
by prolonging or shortening, respectively, the persistence of
quantum features in the initial state of the system. Moreover, we
suggest a physical context in which the Zeno--anti-Zeno crossover
can be observed with current technology by means of reservoir
engineering techniques \cite{engineerNIST}.

The QBM microscopic Hamiltonian model consists of a quantum
harmonic oscillator linearly coupled to a quantum reservoir
modelled as a collection of non-interacting harmonic oscillators
at thermal equilibrium. In the limit of a continuum of frequencies
$\omega$, the reservoir properties are described by the reservoir
spectral density $J(\omega)$ measuring the microscopic effective
coupling strength between the system oscillator and the
oscillators of the reservoir.

One of the advantages of the QBM model is that, for factorized
initial conditions, it can be described by means of the following
exact master equation \cite{HuPazZhang,EPJRWA,PRAsolanalitica}
\begin{eqnarray}
 \frac{d \rho_S(t)}{dt} &=&  \frac{1}{i\hbar}[H_0, \rho_S(t)]
 -   \Delta(t) [X,[X,\rho_S(t)]]  \nonumber \\
 &+& \Pi(t) [X,[P,\rho_S(t)]]+ \frac{i}{2} r(t) [X^2,\rho_S(t)] \nonumber \\
 &-& i \gamma(t) [X,\{P,\rho_S(t)\}], \label{QBMme}
\end{eqnarray}
where $\rho_S(t)$ is the reduced density matrix, and $H_0 = \hbar
\omega_0 \left( a^{\dag} a + \frac{1}{2} \right)$ is the system
Hamiltonian, with $a$, $a^{\dag}$, and $\omega_0$ the annihilation
operator, the creation operator and the frequency of the system
oscillator, respectively. Moreover $X = \sqrt{\omega_0/2 \hbar}
\left( a + a^{\dag}\right) $ and $P=i/\sqrt{2 \hbar \omega_0}
\left( a^{\dag}- a\right) $ are the system position and momentum
operators, where we have set the mass of the particle $m=1$.

The master equation given by Eq.~(\ref{QBMme}), being exact,
describes also the non-Markovian short time system-reservoir
correlations due to the finite correlation time of the reservoir.
In contrast to other non-Markovian dynamical systems, this master
equation is local in time, i.e.~it does not contain memory
integrals. All the non-Markovian character of the system is
contained in the time dependent coefficients appearing in the
master equation (for the analytic expression of the coefficients
see, e.g., Ref.~\cite{Maniscalco04b}). These coefficients depend
uniquely on the form of the reservoir spectral density. The
coefficient $r(t)$ describes a time dependent frequency shift,
$\gamma(t)$ is the damping coefficient, $\Delta(t)$ and $\Pi(t)$
are the normal and the anomalous diffusion coefficients,
respectively \cite{HuPazZhang}. In the secular approximation,
i.e.~after averaging over the rapidly oscillating terms appearing
in the dynamics, the only two relevant coefficients are the
diffusion coefficient $\Delta(t)$ and the damping coefficient
$\gamma(t)$ \cite{Maniscalco04b}. A simple view of the effect of
the diffusion term can be given by looking at the approximate
solution \cite{HuPazZhang} of the master equation given by
Eq.~(\ref{QBMme}), in the position space,
\begin{equation}
\rho_S(x,x',t)\!\simeq \!\rho_S(x,x',0) \exp \left[ -(x-x')^2
\!\!\!\int_0^t \!\!\!\! \Delta(t_1) dt_1 \right]\!\!.
\label{eq:envinde}
\end{equation}
The previous equation shows that the diffusion term is responsible
for the vanishing of the off-diagonal terms of the density matrix
in position space, i.e.~for EID.

For the sake of concreteness, we focus on the case of an Ohmic
spectral density with Lorentz-Drude cutoff (see \cite{Weissbook},
p.25), $J(\omega)= (\omega / \pi) \omega_c^2 /
(\omega_c^2+\omega^2)$ where $\omega_c$ is the cutoff frequency.
This form of the spectral density is one of the most commonly used
since it leads to a friction force proportional to velocity, which
is typical of dissipative systems in several physical contexts.
The main result of the paper, however, holds for general forms of
the spectral density.

We assume that the system oscillator is initially prepared in one
of the eigenstates of its Hamiltonian, i.e.~a Fock state $\vert n
\rangle $. During the time evolution the system is subjected to a
series of non-selective measurements, i.e. measurements which do
not select the different outcomes \cite{petruccionebook}. We
indicate with $\tau$ the time interval between two successive
measurements, and we assume that $\tau$ is so short (and/or the
coupling so weak) that second order processes may be neglected.
Stated another way we assume that the probability $P_n(t)$ of
finding the system in its initial state $\vert n \rangle$,
i.e.~the survival probability, is such that $P_n(t)\simeq 1 $ and
$P_n(t)\gg P_{n\pm 1}(t)$. After $N$ measurements the survival
probability reads \cite{FacchiPRL01,FacchiRev}
\begin{eqnarray}
P_n^{(N)}(t)= P_n(\tau)^N \equiv \exp\left[ - \gamma_n^Z(\tau)
t\right], \label{ptau2}
\end{eqnarray}
where $t=N \tau$ is the effective duration of the experiment and
the effective decay rate $\gamma_n^Z(\tau)$ is defined by the last
equality. In Eq.~(\ref{ptau2}), we have assumed that the
probability $P_n(\tau)$ factorizes. This assumption is justified
by the fact that, in the following we will use second order
perturbation theory. As shown, e.g., in Ref.~\cite{Lax}, up to
second order in the coupling constant, the density matrices of the
system and of the environment factorize at all times.

The behaviour of the effective decay rate, appearing in
Eq.~(\ref{ptau2}), defines the occurrence of the Zeno or anti-Zeno
effect. We indicate with $\gamma_n^0$ the Markovian decay rate of
the survival probability, as predicted by the Fermi Golden Rule.
If a finite time $\tau^*$ such that
$\gamma_n^Z(\tau^*)=\gamma_n^0$ exists, then  for $\tau < \tau^*$
we have $\gamma_n^Z(\tau)/\gamma_n^0<1$, i.e.~the measurements
hinder the decay (QZE). On the other hand, for $\tau
> \tau^*$ we have $\gamma_n^Z(\tau)/\gamma_n^0>1$ and the measurements
enhance the decay (AZE) \cite{FacchiPRL01}.

For an initial Fock state $\vert n \rangle$ there exist two
possible decay channels associated with the upward and downward
transitions to the states $\vert n+1 \rangle$ and $\vert n-1
\rangle$, respectively. The probability that the oscillator leaves
its initial state after a short time interval $\tau$ can be
written as $\bar{P}_n(\tau)=P_n^{\uparrow} (\tau)  +
P_n^{\downarrow} (\tau) $, where $P_n^{\uparrow}(\tau)$ and
$P_n^{\downarrow} (\tau) $ are the probabilities that an upward or
a downward transition, respectively, has occurred in the interval
of time $0 < t < \tau$. From Eq.~(12) of
Ref.~\cite{Maniscalco04b},  neglecting second order processes, one
gets
\begin{eqnarray}
\frac{ d \rho_{nn} (t)}{dt} \!=\! -
\left\{\!(n\!+\!1)\!\left[\Delta(t)\!-\!\gamma(t) \right] \!+\! n
\left[\Delta(t)\!+\!\gamma(t) \right] \right\}\rho_{nn}(t),
\label{eq:another}
\end{eqnarray}
where $\rho_{nn}(t) = \langle n \vert \rho_S (t) \vert n \rangle$.
From Eq.~(\ref{eq:another}), noting that $\rho_{nn}(t) \simeq
\rho_{nn}(0)$, it is straightforward to see that the upward and
downward transition probabilities in the interval $0 < t < \tau$
are $P_n^{\uparrow} (\tau) = (n+1) \int_0^{\tau} [\Delta (t) -
\gamma(t)] dt$ and $P_n^{\downarrow} (\tau) = n \int_0^\tau
[\Delta (t) +\gamma(t)] dt$, respectively. We note that the
survival probability $P_n(\tau)$ defined in Eq.~(\ref{ptau2}) is
given by $P_n(\tau)=1-\bar{P}_n (\tau)$. Assuming that $\tau$ is
small enough and keeping only the first two terms in the expansion
of the exponential appearing in Eq.~(\ref{ptau2}) one easily gets
\begin{eqnarray}
\gamma_n^Z(\tau) =\frac{1}{\tau}\left[ \left( 2n+1 \right)
\int_0^{\tau} \Delta (t) dt - \int_0^{\tau} \gamma(t) dt \right].
\label{gammanZ}
\end{eqnarray}
We notice that the quantity $\gamma_n^Z(\tau)$ can also be derived
starting from the generalized master equation obtained in
Ref.~\cite{Facchi05}, applying the formalism to the case of the
harmonic oscillator. By definition $\Delta(t)$ and $\gamma(t)$, up
to the second order in the coupling constant, are given by
\cite{HuPazZhang,Maniscalco04b}
\begin{eqnarray}
\!\Delta(t)\!\!\!&=&\!\! \!\alpha^2\!\! \!\!\int_0^t \!\!\!\!
\int_0^{\infty}\!\!\!\!\!\! \!\!J(\!\omega\!)\! \coth\!\!
\left(\!\hbar \omega/2 k_B T \!\right) \!\cos(\omega t_1\!) \!\cos
(\omega_0 t_1\!) d \omega d t_1,
\label{delta} \\
\gamma(t)\!\!&=& \alpha^2 \int_0^t\!\! \int_0^{\infty}\!\!\!
J(\omega) \sin(\omega t_1) \sin (\omega_0 t_1) d \omega d t_1,
\label{gamma}
\end{eqnarray}
with $\alpha$ microscopic dimensionless system-reservoir coupling
constant. We note that, for high reservoir temperatures $T$,
$\Delta(t)\gg \gamma(t)$. Inserting
Eqs.~(\ref{delta})-(\ref{gamma}) into Eq.~(\ref{gammanZ}) and
carrying out the double time integration, one obtains
\begin{eqnarray}
\gamma_n^Z(\tau) \!\!=\!\! \tau  \!\!\int_0^{\infty}
\!\!\!\!\!\!\!J(\omega) \left\{ n^{(-)}_{\omega} {\rm
sinc}^2\!\!\! \left(\omega_-\tau \right) + n^{(+)}_{\omega} {\rm
sinc}^2 \!\!\!\left(\omega_+\tau \right) \right\} d \omega,
\label{gammanZII}
\end{eqnarray}
with ${\rm sinc}(x) = \sin(x)/x$, $\omega_{\pm}= (\omega\pm
\omega_0)/2$, and $n^{(\pm)}_{\omega}=\alpha^2[\coth \left(\hbar
\omega/2 k_B T \right) (n+1/2) \pm 1]$. In the limit $\tau
\rightarrow \infty$ one gets the Markovian value of the effective
decay rate $\gamma_n^0 = (2n+1) \Delta_M + \gamma_M$, with
$\Delta_M$ and $\gamma_M$ Markovian values of the diffusion and
damping coefficients respectively.

The quantity ruling the occurrence of either the QZE or the AZE is
the ratio
\begin{eqnarray}
\frac{\gamma_n^Z (\tau)}{\gamma_n^0}= \frac{\left( 2n+1 \right)
\int_0^{\tau} \Delta (t) dt - \int_0^{\tau} \gamma(t) dt}{\tau
[(2n+1) \Delta_M - \gamma_M] }. \label{gammaratio}
\end{eqnarray}
It is worth underlining that, in general, this quantity depends on
the initial state of the system, i.e. on the initial Fock state
$\vert n \rangle$. For high $T$, however, since $\Delta(t)\gg
\gamma(t)$, Eq.~(\ref{gammaratio}) becomes independent of the
initial state
\begin{eqnarray}
\frac{\gamma_n^Z (\tau)}{\gamma_n^0}&\simeq& \frac{ \int_0^{\tau}
\Delta(t) dt}{\tau  \Delta_M}. \label{eq:gammaz}
\end{eqnarray}
Equation~(\ref{eq:gammaz}) approximates Eq.~(\ref{gammaratio})
also for initial highly excited Fock states, i.e. $n \gg 1$. This
equation, together with Eq.~(\ref{gammaratio}), establishes a
connection between two fundamental phenomena of quantum theory,
namely the quantum Zeno effects and environment induced
decoherence. This connection and its physical consequences, which
will be carefully described in the rest of the paper, constitute
our main result. We underline that, in deriving
Eqs.~(\ref{gammaratio})-(\ref{eq:gammaz}), no assumption on the
form of the spectral density has been done.


\begin{figure}
\includegraphics[width=8 cm,height=8 cm]{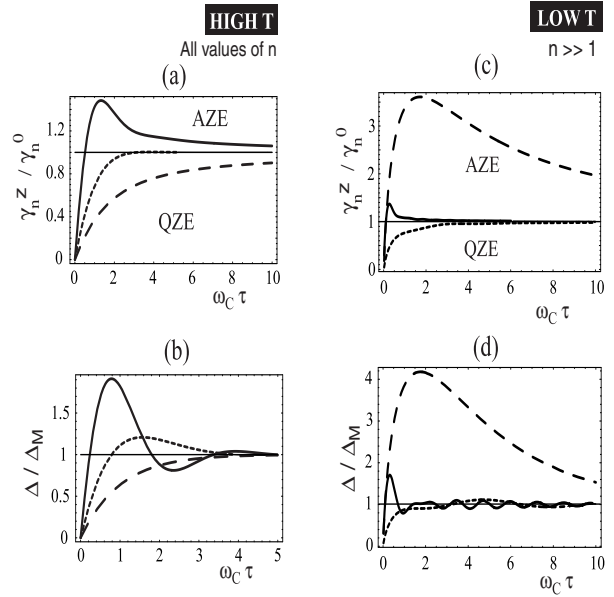}
\caption{\label{fig:1} (a) Ratio between the effective decay rate
$\gamma_n^Z(\tau)$ and the Markovian decay rate $\gamma^n_0$ for
high $T$. Solid line $r=0.5$, dotted line $r=1$, dashed line
$r=10$. (b) Ratio between the diffusion coefficient $\Delta(\tau)$
and its Markovian value $\Delta_M$ for high $T$. Solid line
$r=0.5$, dotted line $r=1$, dashed line $r=10$. (c) Ratio between
the effective decay rate $\gamma_n^Z(\tau)$ and the Markovian
decay rate $\gamma^n_0$ for $T\simeq0$ and $n\gg1$. Solid line
$r=0.1$, dotted line $r=0.5$, dashed line $r=10$. (d) Ratio
between the diffusion coefficient $\Delta(\tau)$ and its Markovian
value $\Delta_M$ for $T\simeq 0$. Solid line $r=0.1$, dotted line
$r=0.5$, dashed line $r=10$. A comparison between (a) and (c)
shows the temperature dependence of the Zeno--anti-Zeno
crossover.}
\end{figure}


Equation (\ref{eq:gammaz}) tells us that, for high $T$ reservoirs
and/or $n \gg 1$ the effective decay rate depends {\it only} on
the diffusion coefficient $\Delta(t)$, which in turn describes
EID. This yields a new physical explanation of the occurrence of
either Zeno or anti-Zeno effects for QBM . The eigenstates of the
quantum harmonic oscillator are highly nonclassical states. They
are not localized, either in position or in momentum, and
therefore they are very sensitive to environment induced
decoherence. If the average EID in the interval between two
measurements, quantified by $\int_0^{\tau} \Delta(t) dt /\tau$, is
smaller than the Markovian one, quantified by $\Delta_M$, then
when the system \lq\lq restarts\rq\rq \; the evolution after each
measurement, the effect of EID is again less than in the Markovian
case. Essentially the measurements force the system to experience
repeatedly an effective EID which is less strong than the
Markovian one. In this case the QZE occurs. On the contrary,
whenever the average increase of EID in the interval between two
measurements is greater than its Markovian value, then the
measurements force the system to experience always a stronger
decoherence, and hence the system decay is accelerated (AZE).

An important consequence of Eq.~(\ref{eq:gammaz}) is that  the
quantum--classical border can be manipulated by means of the QZE
and of the AZE. When the QZE occurs, indeed, an initial
delocalized state of the harmonic oscillator such as its energy
eigenstate remains delocalized for longer times, hence the QZE
\lq\lq moves forward in time\rq\rq \; the quantum--classical
border. The opposite situation happens when the AZE occurs.

The analytic expression of the diffusion coefficient allows us to
single out the relevant system and reservoir parameters ruling the
crossover between the QZE and the AZE. The time $\tau^*$ at which
$\gamma^Z(\tau^*)=\gamma_0$, when it exists and is finite, can be
seen as a transition time between Zeno and anti-Zeno phenomena
\cite{FacchiPRL01}. Our analysis shows that there exist  two other
relevant parameters, namely the ratio $r= \omega_c/\omega_0$,
quantifying the asymmetry of the spectral distribution, and the
ratio $k_B T/ \hbar\omega_0$. For a fixed time interval $\tau$, a
change in the values of $r$ and $T$ may lead to a passage from
Zeno dynamics to anti-Zeno dynamics and vice versa.

For high $T$,  it is easy to prove that a Zeno--anti-Zeno
crossover exists only for values of $r < 1$, i.e.~for
$\omega_0>\omega_c$, as it is shown in Fig \ref{fig:1} (a). This
can be understood in terms of EID by looking at the short time
dynamics of $\Delta(t)$ in the time interval $0<t<\tau$,
i.e.~before the first measurement is performed (see
Fig.~\ref{fig:1} (b)).

The situation changes drastically for the case of a zero-$T$
reservoir, characterized by an asymmetric spectral density. For $n
\gg 1 $, indeed, in contrast to the high $T$ case, the
Zeno--anti-Zeno crossover exists also for values of $r\gg 1$.
i.e.~in the case $\omega_0 \ll \omega_c$. The region in which only
Zeno dynamics may occur now appears at the edge of the spectral
density function, i.e.~for $\omega_0 \simeq \omega_c$. The reason
of such a different dynamics for small values of $\omega_0$ stems
from the strong decoherence, showing up at low temperatures not
only when $r \ll 1$ (as in the high $T$ case), but also for $r \gg
1$ (See Figs.~\ref{fig:1} (c)-(d)). Summarizing, the occurrence of
the Zeno or anti-Zeno effects is directly related to the absence
or presence, respectively, of the so-called initial jolt of the
diffusion coefficient $\Delta(t)$ \cite{HuPazZhang}, which is the
signal of strong initial decoherence.

Another interesting aspect stemming from our results concerns the
Zeno-anti--Zeno crossover for an initial ground state ($n=0$) of
the system oscillator. The high $T$ behaviour is given by
Eq.~(\ref{eq:gammaz}) [See Fig.~\ref{fig:1} (a)-(b)]. For very low
reservoir temperatures ($T \simeq 0$), $\Delta_M \simeq \gamma_M$
and therefore the denominator of Eq.~(\ref{gammaratio}) approaches
zero, implying that $\gamma^{Z}(\tau)/\gamma_0 \gg 1$ always, i.e.
the measurements always enhance the decay (AZE). Summarizing, in
this case, by changing the reservoir temperature, e.g. starting
from high $T$ and lowering the temperature, one observes a passage
from the situation depicted in Fig. \ref{fig:1} (a) to a situation
in which only the AZE is practically observable.

The possibility of controlling both the environment and the
system-environment coupling would allow one to monitor the
transition from Zeno to anti-Zeno dynamics. The  use of artificial
controllable engineered environments has been recently
demonstrated for single trapped ions \cite{engineerNIST}. In
Ref.~\cite{Maniscalco04a} it has been shown that the engineered
amplitude reservoir realized by applying noisy electric fields to
the trap electrodes in \cite{engineerNIST} can be used to simulate
quantum Brownian motion and to reveal the quadratic short time
dynamics. Shuttering these noisy electric fields  one could model
a fast switch off--on of the environment. Actually, when the noise
is off, the reservoir simply does not exist anymore. The action of
the sudden switch off-on of the environment may be seen as a
physical implementation of the operation of trace over the
reservoir degrees of freedom.   The operation of trace is a
typical example of a non-selective measurement (see, e.g.,
Ref.~\cite{petruccionebook}, p. 321). Hence a succession of short
switch off-on periods, realized by shuttering the engineered
applied noise, would induce Zeno or anti-Zeno dynamics depending
on the value of the system and reservoir parameters and of the
shuttering period. This is the core idea for monitoring the
Zeno--anti-Zeno crossover with trapped ions.

It is worth underlining that since the measurements causing the
acceleration or inhibition of the decay, implemented by a fast
switch off-on of the noisy electric field, are { \it
non-selective}, they do not disturb the vibrational state of the
ion. Therefore, by using the set up of Ref. \cite{engineerNIST}, a
comparison between the population $P_n^{(N)}(t)$ of the initial
vibrational state (e.g. the vibrational ground state $\vert n=0
\rangle$)  in presence of {\it shuttered noise} (with $N$ the
number of switching off-on periods), and the population $P_n(t)$
 in presence of {\it un-shuttered noise}, would show that
$P_n^{(N)}(t)>P_n(t)$ (QZE) or $P_n^{(N)}(t)< P_n(t)$ (AZE)
depending on the choice of the parameters.

All the parameters $\omega_0$, $\omega_c$, $T$ and $\tau$, driving
the Zeno--anti-Zeno crossover, may be varied in the experiments.
Although the value of $\omega_0$ may be modified only within a
certain range and under certain constrains, the modification of
both $\omega_c$ and $T$ may be obtained by simply filtering the
applied noise and varying the noise fluctuations, respectively
\cite{Maniscalco04a}. Since all the parameters ruling the Zeno --
anti-Zeno crossover are easily adjustable in the experiments, its
observation in the trapped ion context should be already in the
grasp of the experimentalists.

The authors acknowledge financial support from the Academy of
Finland (projects 207614, 206108, 108699,105740), the Magnus
Ehrnrooth Foundation, and the EU's project CAMEL (Grant No.
MTKD-CT-2004-014427). Stimulating discussions with David Wineland,
Paolo Facchi and Saverio Pascazio are gratefully acknowledged.


\end{document}